\documentclass{article}

\usepackage{arxiv}

\usepackage[utf8]{inputenc} 
\usepackage{hyperref}       
\usepackage{url}            
\usepackage{booktabs}       
\usepackage{amsfonts}       
\usepackage{nicefrac}       
\usepackage{microtype}      
\usepackage{lipsum}

\usepackage{natbib}
	\setcitestyle{authoryear,round,aysep={}}
\usepackage{nameref}
\usepackage[figuresleft]{rotating}
\usepackage{enumitem} 
\usepackage{amsmath}
\usepackage{graphicx}
\usepackage{algorithm}
\usepackage{algorithmicx}
\usepackage{algpseudocode}
\usepackage[normalem]{ulem} 
\useunder{\uline}{\ul}{} 
\usepackage{xcolor}
\usepackage{subcaption}

\usepackage[noblocks]{authblk}

\newenvironment{itemize*}{\begin{itemize}}{\end{itemize}}
\newenvironment{enumerate*}{\begin{enumerate}}{\end{enumerate}}
\newenvironment{description*}{\begin{description}}{\end{description}}
\newcommand{\email}[1]{\def \@email{#1}}
\usepackage{endnotes}
\renewcommand{\footnotesize}{\normalsize}
\newcommand{\jassonly}[1]{}

\newcommand*\justify{%
  \fontdimen2\font=0.4em
  \fontdimen3\font=0.2em
  \fontdimen4\font=0.1em
  \fontdimen7\font=0.1em
  \hyphenchar\font=`\-
}
\newcommand{\umltext}[1]{\texttt{\justify #1}}
\newcommand{\namerefopt}[1]{\ref{#1}}
\newcounter{nameOfYourChoice}
\graphicspath{{./Figures/}}

\title{Modelling Agents Endowed with Social Practices: Static Aspects}
\author[1]{Rijk Mercuur}
\affil[1]{Faculty of Technology, Policy and Management, Delft University of Technology}
\affil[2]{Faculty of Electrical Engineering, Mathematics and Computer Science, Delft University of Technology}
\affil[3]{Leiden Institute of Advanced Computer Science, Leiden University}

\author[1]{Virginia Dignum}
\author[2,3]{Catholijn M. Jonker}

\begin{document}
\maketitle

\begin{abstract}
To understand societal phenomena through simulation, we need computational variants of socio-cognitive theories. Social Practice Theory has provided a unique understanding of social phenomena regarding the routinized, social and interconnected aspects of behaviour. This paper provides the Social Practice Agent (SoPrA) model that enables the use of Social Practice Theory (SPT) for agent-based simulations. We extract requirements from SPT, construct a computational model in the Unified Modelling Language, verify its implementation in Netlogo and Prot\'eg\'e and show how SoPrA maps on a use case of commuting.  The next step is to model the dynamic aspect of SPT and validate SoPrA's ability to provide understanding in different scenario's. This paper provides the groundwork with a computational model that is a correct depiction of SPT, computational feasible and can be directly mapped to the habitual, social and interconnected aspects of a target scenario. 
\end{abstract}

\keywords{
Social Practice Theory \and Social Agents \and Habits \and Routines \and Social Intelligence \and Interconnected Actions \and Activity Hierarchies
}

\jassonly{\parano{}}


\section{Introduction}
To understand societal phenomena through simulation, we need computational variants of socio-cognitive theories. This paper provides the Social Practice Agent (SoPrA) model that enables the use of Social Practice Theory (SPT)  for agent-based simulations. SPT studies our daily ‘doings and sayings’, such as commuting, dining or greeting. SPT focusses on attributes of activities, namely related values, activities, agents, resources or locations. SPT studies the similarity of these attributes over time (addressing habits), between people (addressing social intelligence) and between different actions (addressing interconnected behaviour). SoPrA uses this perspective to increase our understanding of social phenomena.

SPT has provided a unique understanding of social phenomena in sociology. \cite{Giddens1984} showed how SPT captures that our actions are part of a routine and often not conscious, voluntary or intentional. \citet{Reckwitz2002} explained how SPT captures that the social world is formed and shaped by shared practices and not an additional layer of joint intentions or conventions. \citet{Shove2012} showed how our life is full of bundles of social practices: interconnected actions that are similar and influence each other and that cannot be seen in isolation. By using these aspects of SPT in social simulation we can provide potential explanations that increase our understanding of social phenomena \citep{Grabner2018}. Although there are several useful conceptual frameworks and implementations \citep{Holtz2014,Narasimhan2017,DignumSP}, there is no computational model of SPT that can explain social phenomena in terms of the habitual, social and interconnected aspects of a SP.

To show that SPT can provide potential explanations through simulation we need a computational model that is validated. This paper focuses on the pre-requisite steps to validation: constructing a computational model that is a correct depiction of SPT, computational feasible and clearly maps the model onto a target scenario \citep{Grabner2018}. We approach this by surveying the literature on SPT and summarizing the core aspects of SPT in requirements. We construct a computational model in the Unified Modelling Language (UML) that fulfills these requirements and implement the model in both Netlogo \citep{Netlogo} and Prot\'eg\'e \citep{Protege}. By showing how our UML fulfills the requirements extracted from SPT and, in turn, our implementations correctly depict our UML, we verify that we made a computational model that is both feasible and a correct depiction of SPT. Modelling choices throughout this process are communicated and aimed at the computational feasibility or the correct depiction of SPT. A use case on the societal problem of CO-2 emission due to car commuting shows how SoPrA clearly maps onto habitual, social and interconnected aspects of a social problem. Future work is to model the dynamic aspect of SPT and validate SoPrA. The computational model provided can be used to provide potential explanations of social situations or, possibly outside social simulation, to guide human-agent interaction in a socially aptitude and efficient way.

The remainder of the paper is structured as follows. Section \namerefopt{sec:usecase} explains the scenario that will be used as a use case throughout the paper. Section \namerefopt{sec:literature} captures SPT in requirements and uses supporting literature from psychology and agent theory to correctly model the core aspects of SPT. Section \namerefopt{sec:sopra:highlevel} and \namerefopt{sec:sopra.uml} present a high level description and UML version of the SoPra Model. Section \namerefopt{sec:usecase2} shows how we can make a clear mapping from SoPrA to the habitual, social and interconnected aspects of our scenario.

\section{Use Case \& Descriptive Statements}
\label{sec:usecase}
This section describes the use case of a modeller who aims to explain the societal problem of CO-2 emission due to car commuting. The core mechanisms of SPT are essential in car commuting. First, social intelligence is used when people coordinate their commute: carpool, bring their kids to school. Second, a meta-study found that habit is the strongest predictors of car use; stronger than other frequently studied cognitive measures such as attitude, norm or values \citep{Hoffmann2017}.  Third, \citet{Cass2016} found in a series of interviews that an important consideration in taking the car is the interconnectedness of the work-commute, sport-commute and school-commute. For example, some of the interviewed took the car to work, because this allowed them to combine that action with bringing their kids to school. The following describes a scenario of how Bob can come to the decision to take the train instead of the car.

\begin{quote}
Bob and Alice are at home. Bob usually takes the car to brings the kids to school and go to work. He recently started to value the environment more, but does not really reflect on this habit when at home with his kids. Furthermore, he equates commuting with car commuting. For him, they are alike. Today the kids are not home. Alice asks Bob if he wants to go by train with her? She beliefs Bob wants to commute, that biking is not very comfortable with the rain outside and that people that go by car usually value comfort. She beliefs that normally when others brings the kids to school by car, they will also go to work by car. However, now that the situation is different, he will reconsider. Bob agrees to go by train. He does not have a habit in this situation and that activity promotes environmentalism.
\end{quote}
Section \namerefopt{sec:usecase2} shows how SoPrA can be used to capture this scenario.

\section{Literature and Requirements}
\label{sec:literature}
This section describes related literature and empirical work on SPT (Section \namerefopt{sec:literature:SPT}) and on agent and psychology literature that explores key properties of SP (Section \namerefopt{sec:literature:key}).  We end each section with a list of requirements. These requirements summarize the section and extract what we find most important. 
Section \namerefopt{sec:sopra.uml} presents a UML model that meets these requirements. 

\subsection{Social Practice Theory}
\label{sec:literature:SPT}

\subsubsection{Short History of Social Practice Theory}
Social practice theory has its roots in the work of Wittgenstein and Heidegger who both ascribe a central role to \emph{praxis} or practices. This starts a tradition where the central unit with which one describes the world are our daily doings (e.g., breakfast, work, meetings, commuting). This is in contrast with the traditional view in sociology to see the world in terms of agency (e.g., an employee) and social structures (e.g., an organization). For \citet{Giddens1984} in particular, practices are a way to bridge these two levels of abstraction as a practice exist on both levels (e.g., a meeting influences the employee and the organization). \citet{Bourdieau1977} and \citet{Schatzki1996} popularized the concept of practices in the 1980s and 1990s, where Bourdiea focussed on the practice as a personal `habit', Schatzki focussed on the practice as something social. \citet{Reckwitz2002} further emphasized the importance of SPs as a new way to capture the social world that breaks with old paradigms: the social is formed and shaped by shared practices and not an additional layer of joint intentions or conventions.  Currently, the most common conceptualization of SP comes from \citet{Shove2012} who sees the concept of practices as a collection of ever changing interconnected elements. For example, the practice of driving is a connection of the material car, the meaning of transport and the competence to drive it. Recently, \citet{DignumSP} used SP as a parsimonious concept around which one can order relevant social knowledge for an agent. They divert from the original line by bringing agency back on the table and linking SP to standard agent concepts such as goals, norms and beliefs. SPT thus started as a focus on `doing' and continued to conceptualize that `doing' as routines, that are shared between multiple people and relate different elements. 

\subsubsection{Key Properties of Social Practice} We focus on three key properties of SPs that reflect three aspects of human behaviour: habits, sociality and  interconnected behaviour. These key properties provide an aim and scope for our computational agent model of SP.

SPs are first of all practices. They are habitual enacted routines \citep{Shove2012,Reckwitz2002,Dignum}. Such routines are repeated over time and exist by virtue of this reproduction \citep{Shove2012}. \citet{Reckwitz2002} emphasized that this routine is both physical and mental: there is repetitiveness in both the bodily movement and mental elements that are associated with the SPs. 

SPs are social. \citet{Shove2012} and \citet{Reckwitz2002} interpret social here as that the SP is similar for multiple people. For example, most people share the belief that one needs the know-how to control a car to do the SP of car driving. \citet{Reckwitz2002} sees SPs as a new way to capture the social world that breaks with old paradigms: the social is captured in our shared SPs instead of in a shared mental world (e.g., as described by BDI-agents), in texts (e.g.,in scientific papers) or interactions (e.g., in a discussion). \citet{Reckwitz2002} this similarity means that the SP is something collective, ``but not in the sense of a mere sum of the content of single minds", but in a time-space transcending non-subjective way. We assume that \citet{Reckwitz2002} refers here to the SP as something macro. A separate entity that emerges out of micro enactments, but is more than just the sum of them. Note that common knowledge dictates that SPs are often not fully shared (i.e., similar for all people). For example, Bob and Alice might share many beliefs about the SP of commuting, but Alice associate car-driving with efficiency and Bob with fun. As becomes clear in the next section, modelling this tension between the individuality and sharedness of a SP is one of the main challenges this paper tackles. 

There is another interpretation of sociality in SP: sociality as interaction. This interpretation becomes clear when \citet{DignumSP} zoom in on the micro aspects of SPs. For example, greeting, meeting, discussing and teaching are all SPs that can only be enacted by multiple people. This meaning of sociality is closer to the layman perspective on a social activity: an activity done in the presence of other people. SPs having a guiding role in this interaction, allowing one to form expectations about the actions of others.
This view contrasts with the view of \citet{Shove2012} and \citet{Reckwitz2002}, who deem a SP social if its shared, even if there is no interaction when enacting it. For example, one of \citet{Shove2012}'s canonical examples of a SP is showering, a SP mostly done alone. 
We claim these two interpretations of sociality are reciprocally related. First, a SP only becomes shared through interaction. In the case of greeting, the SP becomes shared through direct interaction. You see how others greet and adapt your ideas of a SP. In the case of showering, the SP becomes shared through indirect interactions (judging remarks, sighs, sniffles). Second, a SP can guide the interaction because it is shared. One forms expectations of the actions of the other based on the assumption that the actions, meanings, habits (i.e., elements of a SP) are shared. 

Different practices are similar and influence each other. \citet{Shove2012} called such similar practices: bundles of practices that are interconnected. They argued that if practices are similar in some aspects, then they become similar in other aspects too. We argue that SPs are interconnected in terms of common elements, time and space. First, SPs connect when they are enacted at the same time or in sequence. For example, in a series of interviews, \citet{Cass2016} found that one main reason people take the car to work is because this makes it easier to combine this SP with SPs of leisure, healthcare or shopping. Second, SPs connect when they are enacted in the same space. For example, a supermarket at the train station facilitates the connection of the practice of groceries and commuting. Third, SPs might be connected, because they share an element.  For example, \citet{Shove2012} mentions how in the early days of driving, cars easily broke down. To be able to drive a car one had to have the competence to repair it. The practice of driving thus became connected with other practice that related to the competence of repairing, for instance, plumbing or carpeting. The meaning of these practices as something masculine influenced the meaning of car driving. Car driving and plumbing now share this element of meaning as something primarily masculine.

\begin{enumerate*}[label=\textbf{SPT.\arabic*},align=left]
\item SPs are habitually enacted routines. \label{req:SPT.habits}
\item SPs are social:
\begin{enumerate}
\item SPs are similar for multiple agents. \label{req:SPT.socialsimilar}
\item SPs influence and are influenced by agent interactions.
\end{enumerate}
\item SPs are interconnected: in terms of time, space and common elements.\label{req:SPT.interconnected}
 \setcounter{nameOfYourChoice}{\value{enumi}}
\end{enumerate*}

\subsubsection{The elements of a Social Practice}
What comprises a SP? Although short names, such as "greeting" are given to SPs, SPs are collections of interconnected elements. This idea started with \citet{Taylor1973} for who "meanings and norms implicit in [...] practices are not just in the minds of the actors but are out there in the practices themselves". For Taylor a practice is not merely a set of actions, but also comprises other concepts. For \citep{Reckwitz2002} this means the SP's existence necessarily depends on the existence and specific interconnectedness of these elements, but cannot be reduced to any single one of them. The practice of skateboarding thus only exists because a group of elements comes together: materials like the skateboard, street spaces, the competences to ride the board, rules and norms that define the practice and the meaning different groups attribute to the practice \citep{Shove2012}. For example, as a recreational activity, an art form or a means of transport. We continue by describing the literature on those elements we think are necessary and sufficient to model the key properties.

Although A SP is thus not merely (a set of) action(s), an action is a commonly agreed aspect of a SP. Actions here refer to the bodily movement itself without all the mental connotations. We discern between two views in the literature. First, according to \citet{Reckwitz2002} a SP represents a pattern which can be filled out by a multitude of single and often unique actions reproducing the SP. For example, the SP of commuting can consist of taking your kids to school and then going to work.  Second, for \citet{Shove2012} actions are not a part, but instead another view on the SP. She views the SP as either a collection of elements or as something that is performed. For example, commuting can be seen as either a collection of elements such as, the meaning of transport and the material car or as a series of actions, such as, getting in your car and going to work. We follow \citet{Reckwitz2002} as (1) it allows us to retain ourselves to the concept of 'elements' (instead of introducing an auxiliary concept 'perspective') in line with our aim for parsimony (2) modelling SPs as a hierarchy of actions allows us to link SPT to agent models that focus on actions as well.

There are symbolic and social elements associated with a SP, such as, the symbolic element 'health' to the practice of riding the bike. These symbolic and social associations are conceptualized differently by different authors. \citet{Schatzki1996} emphasized the 'teleoaffective' social elements, such as, embracing ends, purposes, projects and emotions. He described how snapshots of what people do have a history and a future that are united in the moment of performance. In our words, actions connect the now with past mental connotations and determine future connotations. \citet{Reckwitz2002} focused on mental, emotional and motivational knowledge. For Reckwitz it's important that the mental is part of the routinized SP. We could not imagine football without the associated aim (to win the game), understanding (of others players behaviour) and emotions (a particular tension). \citet{Shove2012} simplified these concepts in a bucket term called meaning. This has the advantage that one can easily refer to all symbolic and social element of a SP, but the disadvantage that the term is not well-defined or measurable. (In fact, \citet{DignumSP} unfolds this concept of meaning again and differs between social interpretation, roles or norms.) We follow \citet{Shove2012} here for the sake of simplicity: it allows us to model the key properties, cost less time to work out and serves computationally feasibility. Given the key properties we aim to model we highlight two aspects of the meaning element. First, the element of meaning has a teleological role: it represents an end or purpose. Second, this end or purpose is social: its shared by others and has a role in guiding interaction. 


Whereas in the past SPT focussed on these classical social aspects (meanings, norms), more recently \citet{Schatzki1996,Latour1996,Reckwitz2002,Shove2012} have focussed on the physical aspects of a SP. They introduce an element called `materials', which helps in closing the `gap' between the physical and social world \citep{Shove2012}. By abstracting over the actor, SPT shows that for multiple people the same SPs are connected to the same type of physical objects. For example, almost everyone associates a car, road and gas station with the SP of driving. These physical objects play a central role in our social world in that it provides a shared common ground \citep{Searle1995}. \citet{DignumSP} connects the physical dimension of a SP to the decision making of an agent by emphasizing that certain materials can trigger a SP. If the physical context triggers a SP, it means the agent will typically do the SP. As we will show in the next section, triggering a SP is closely connected to  what social psychologist call a habit. 


In sum, this paper focuses on the following elements of a SP: an action element, a meaning element and a physical element. We argue that this set of components is necessary and sufficient to model the aforementioned key properties of a SP. There are more elements that can be routinized, socialized or interconnected. Other commonly described elements of SP are emotions, competences, affordances, norms, goals and plans. Our view is that these are valuable concepts that could improve SoPrA in its ability to model behaviour. However, to show how the mechanisms of routines, sociality and interconnectedness in SP these elements are necessary and sufficient. After introducing SoPrA, we will show how SoPrA can be clearly mapped on this aspects of our use case \namerefopt{sec:usecase2}.

\begin{enumerate*}[label=\textbf{SPT.\arabic*},align=left]
\setcounter{enumi}{\value{nameOfYourChoice}}
\item SPs consists of actions that represent the bodily movement. \label{req:SP.actions}
\item SPs consist of a meaning element that represents a shared end or purpose. \label{req:SPT.meaning}
\item SPs consist of a physical element that \label{req:SPT.context}
\begin{enumerate*}
\item can trigger an action 
\item relates the physical world to the social. 
\end{enumerate*}

\end{enumerate*}

\subsection{Literature on Key Properties}
\label{sec:literature:key}
\subsubsection{Habits}
A key aspect of SPs is that they are repeated over time and exist by virtue of this reproduction~\citep{Reckwitz2002}. Repetitive behaviour is also studied from the individual perspective by social psychologist as 'habits'. They use the term ‘habit’ to refer to a phenomenon whereby behaviour persists because it has become an automatic response to particular, regularly encountered, context \citep{Kurz2014}. The model presented in this paper links the notion of habitual behaviour to the notion of SP. 

Habits are automatic. \citet{Moors2006} compares several theoretical views and concludes that automaticity entails unintentional, uncontrollable, goal independent, autonomous, purely stimulus driven, unconscious, efficient, and fast behaviour. Habits are thus contrasted with predictors of intentional actions, such as values, motives and goals. When a context triggers the habit these intentional concepts do not mediate the relation.\endnote{There is debate about the exact relation between habitual and intentional actions. \citet{Gardner2011} shows in a meta-analysis that in 8 out of 21 studies habits moderate the intention-action relation.}  For example, even if one values the environment higher than efficiency one habitually takes the car to work. Habits can be measured with the Self-Reported Habit Index, a 12-item index that includes items such as `I do frequently' and `I do without thinking' \citep{Verplanken2003}. This measure has a medium-to-strong ($r_+=0.44$) relation with behaviour \citep{Gardner2011}. Habits thus provide a fast automatic heuristic for action that derives its meaning from being contrasted to a slow intentional mode of action. 

Habits are context-dependent~\citep{Kurz2014}. For example, a habit at home might not be a habit at work. One might drink coffee every two hours at work, but not at home. The immediate setting, or context, of an agent thus triggers an action. The literature on habits differs in what they consider as context. A narrow definition focusses on physical tangible elements. For example, nearby cigarettes can trigger a habit of smoking. Wider definitions of context allow time, other activities \citep{Verplanken2005}, location and/or other people to trigger habits. \citet{Wood2007} nuance the context-action relation by conceptualizing habits as `context cue'-action relations. For example, it is not so much work that triggers drinking coffee, but a combination of cue's at work such as a colleague and the coffee machine. The strength of these `context cue'-action relations is continuous instead of discrete (e.g., a coffee machine is a slightly stronger habitual trigger than a colleague) \citep{Moors2006}.

The actor influences the habit. \citet{Lally2010} studied how habits are formed in the real world. Subjects were asked to do the same action daily in the same context and report on automaticity (calculated using a sub-set of SRHI items). She found that humans differ in the maximum automaticity they reported and also in the time to reach this maximum. 

There are two aspects of habits that we consider out of scope of this paper. Different contexts can be similar and thus trigger similar habits. \citet{Mercuur2015} shows that although there is little known about how habits in one context relate to habits in another context this is crucial to model habits in realistic settings. To our knowledge, there is no theoretical or empirical work on this topic. This topic demands a study in itself and is postponed to future work. 
Habits are related to the concept of attention \citep{Wood2007}. The more attention an individual can muster the lower the chance he or she falls into a habit. Attention becomes important when modelling habits over time. Different levels of attention over time give rise to different decisions. As this paper focuses on static aspects this topic is postponed to future work. 


\begin{enumerate*}[label=\textbf{H.\arabic*}, align=left]
\item Habits provide a fast automatic heuristic for action that derives its meaning from being contrasted to a slow intentional mode of action. \label{req:H.automatic}
\item Habits exist by virtue of continuous action-'context cue' associations. \label{req:H.associations}
\item Context cue's that trigger habits include physical elements and could extent to concepts such as time, other activities, location or other people.\label{req:H.contextcue}
\item Humans differ in 
\begin{enumerate*}
\item the height of the maximum automaticity (or strength of the habit) they experience \label{req:H.diffmaximum}
\item the time to reach this maximum \label{req:H.difflearningtime}
\end{enumerate*}
\end{enumerate*}





\subsubsection{Social Intelligence}
The second key aspect of SPs is that they are social. The social world is also studied from the individual perspective as social intelligence. Social intelligence is our ability to act wisely in in a world that we share with others \citep{Thorndike}. The model presented in this paper links the notion of social intelligence to the notion of SP.

As mentioned, sociality in SPT has two faces: sociality as a shared world or sociality as interaction. Social intelligence can also be seen from these two perspectives.

\citet{Castelfranchi1997} describes social intelligence from the first perspective: as our ability to reason in a shared world. The agent literature provides different answers on what this shared world consists of. 
A first series of papers uses agents that only take into consideration the actions of other. For example, in the Consumat model \citep{Jager2000} an agent takes into consideration what most other agents do. \citet{Castelfranchi1997} emphasized that we need to extend such models to also take the mental state of other agents into consideration. The notion of social action cannot be a behavioural notion - just based on an external description, because what makes the action social is that it is based on certain mental states. 
A second series of papers focusses on such a representation of the mind of other agents based on individual notions such as beliefs, desires and intentions \citep{Dignum2000, Felli2014}. Sociality is then introduced as a secondary notion. For example, as the ability to form beliefs about others goals \citep{Felli2014} or as a mechanism to filter its intention to a socially desired set \citep{Dignum2000}. \citet{Dignum2014,Hofstede2017} argued that humans are at the core social beings and thus use social concepts as a primary concept. We view social concepts here as referring to reasoning concepts that depend on being shared. For example, a notion of culture is hard to imagine without multiple agents. 
A third series of paper focuses on these social concepts. For example, values, norms \citep{Dignum1999}, trust \citep{Jonker1999}, culture and identity. What characterizes this work is to bring a collective (social) notion back to the reasoning of the individual. This paper follows this last view in using social concepts as core mechanisms for agents and the basis for our social world.

From the perspective of the individual, one beliefs in a shared social world, but also a personal view. For example, one beliefs that car-driving is usually seen as a means for transport, but beliefs him or herself its a fun activity. These personal and shared aspects of a view are independent: one can belief in a personal view on something without believing this view is shared or one can belief something is the shared view without believing in this view. In other words, there are multiple views on the world that one can all to a certain extent agree with or beliefs others agree with.

Theory of mind considers such reasoning about beliefs of others \citep{Woodruff1978}. In addition, it treats two more aspects: beliefs about specific others and chains of beliefs. For example, one can belief John beliefs car-driving is fun or belief that John beliefs that I believe that John believes car driving is fun. These aspects are out of the scope of SoPrA. SP are a heuristic that  considers only two agents: itself or the group. For example, when greeting someone in most cases it suffices to know that most people view greeting as polite and see shaking hands as a part of greeting. SP focuses on the social intelligence that works in most situation, in contrast, theory of mind treats particular cases where more in-depth reasoning is needed. For a model of SP, it's thus sufficient to require two aspects of our view on the shared social world: to what extent one beliefs this view him or herself and to what extent one beliefs others hold this view.

\citet{Goleman2011} describes social intelligence from the second perspective: as our ability to be aware and reason about this common ground in interactions. He divides social intelligence in social awareness and social influence. Social awareness allows one to form exceptions of other people and social influence to act upon those. As this paper focuses on the static aspects of SPs we postpone the dynamics of interactions to future work.

\begin{enumerate*}[label=\textbf{SI.\arabic*}, align=left]
\item Social intelligence gives the ability to represent and reason about the shared world between agents. \label{req:SI.sharedworld}
\item This shared world needs to be captured in terms of actions, individual concepts and foremost social concepts, that is, concepts that depend on being shared among multiple humans. \label{req:SI.socialbuildingblocks}
\item There are multiple views possible on this shared social world. \label{req:SI.multipleviews}
\begin{enumerate}
\item Agents can belief to a certain extent themselves in a particular view. \label{req:SI.personalview}
\item Agents can belief that others to a certain extent belief in a particular view. \label{req:SI.othersview}
\item These two beliefs can be independent: one can belief something without believing others do and vice versa. \label{req:SI.independentviews}
\end{enumerate}
\item This common ground forms the basis of wise social interactions by allowing social awareness and social influence.  \label{req.socialinteractions}
\end{enumerate*}


\subsubsection{Interconnected Actions}
The third key aspect of SPs is that they are interconnected. 
This subsection uses the work on interconnected actions to model this key aspect. 

In neuroscience, \citet{Metzinger2003} brought together empirical evidence that shows that humans use actions as a basic building block to represent the world. Actions here are the bodily movements conceptualized in the motor system. \citet{Metzinger2003} shows how actions are used by the remainder of the brain as basic constituents of the world it interprets. Humans form ontologies of how these actions are related to each other. \citet{Metzinger2003} further shows how these action ontologies serve as a building block for social cognition: action-associations abstract over the subject, therefore they can be used to map on a first-person or second-person perspective. 

Action ontologies have been studied in the context of smart homes to be able to recognize activities \citep{Kasteren2008,Storf2009,Chen2012,Okeya2014}. \citet{Okeya2014} makes a difference between actions, simple activities and composite activities. Actions are atomic. A simple activity is an ordered sequence of actions. Composite activities are multiple actions within a time interval, but with no strict order to the actions and can thus overlap. For example, driving might consist of overlaps of the activity of taking a turn and listening to the navigation. From this point on we will separate between activities and actions. Activities refer to any bodily movement (i.e., actions, simple activities and composite activities). Actions to refer to the subset of activities that is atomic. 

\citet{Okeya2014} separate two types of relations between activities: an ontological and temporal relation.\endnote{Note that other authors use the term ontology to refer to any kind of relation between two objects. Temporal relations are thus a subset of ontological relations. However, \citet{Okeya2014} uses the term ontological to refer to inferences one can easily make in description logic, whereas he uses the term temporal to refer to relations he can make in Allan's temporal logic.} The ontological part describes relations between actions such as subsumptions, equivalence or disjointness. For example, taking the train to school is a kind of commuting. A temporal relation
encodes qualitative information regarding time. For example, the user performs two activities after another. Humans are naturally able to make ontological and temporal inferences about activities. For example, humans infer that if taking the car to work is environmentally unfriendly then taking the car to school might be as well. Or they might induce one is doing the activity of commuting as one is doing several actions in a sequence: walking to the car, going in the the car, driving to work and working.

In the area of agent communication, interconnected actions have been studied under the name of conversation protocols \citep{Kumar2002}. A central aim in this work is to coordinate the agents by identifying important shared fixed points (called landmarks), while allowing the agents freedom in their path to reach these states. The result of this is an activity hierarchy with partial ordered landmarks that represents the conversation protocol.

\begin{enumerate}[label=\textbf{IA.\arabic*}, align=left]
\item Actions are basic building blocks of how agents represent the world. \label{req:IA.basicbuildingblock}
\item Agents can relate these actions into meaningful activity hierarchies. \label{req:IA.hierarchy}
\item Activity hierarchies serve as a building block for social cognition.
\item Activities can consists of multiple other activities:
\begin{enumerate}
\item sequential activities \label{req.IA.sequential}
\item overlapping activities. \label{req.IA.overlapping}
\end{enumerate}
\item Activity hierarchies have ontological and temporal aspects. \label{req:IA.ontologicaltemporal}
\item Agents are able to make inference about these ontological and temporal properties. \label{req:IA.inferences}
\item Activity hierarchies have shared fixed points, but also allow the freedom to have personal routes towards those points.
\end{enumerate}

\subsection{Overview of Requirements}
Table \ref{table:overview} provides an overview of the requirements extracted from the literature. The requirements are simplified for display purposes.
\begin{table}[h]
\centering
\caption{An overview of the requirements extracted from the literature.}
\label{table:overview}
\resizebox{\textwidth}{!}{%
\begin{tabular}{lllll}
\hline
 & \textbf{Social Practices} & \textbf{Habits} & \textbf{Social Intelligence} & \textbf{Interconnected Activities} \\ \hline
\textbf{1} & are habitual & are fast and not slow & \begin{tabular}[c]{@{}l@{}}gives the ability to reason \\ in a shared world\end{tabular} & are basic building blocks \\
\textbf{2} & are social: similar and interactive & are action- context cue links & \begin{tabular}[c]{@{}l@{}}requires shared actions, \\ individual and social mental concepts\end{tabular} & form activity hierarchies \\
\textbf{3} & are interconnected & triggered by resources, time, etc. & \begin{tabular}[c]{@{}l@{}}requires an independent \\ personal and shared view\end{tabular} & form the basis for social cognition \\
\textbf{4} & consists of actions & different per human & \begin{tabular}[c]{@{}l@{}}provides the basis for social \\ awareness and social influence\end{tabular} & consist of other activities \\
\textbf{5} & consists of a meaning element &  &  & can relate ontologically or temporally \\
\textbf{6} & consists of a physical element &  &  & allows agents to make inferences \\
\textbf{7} &  &  &  & are fixed points or personal deviations \\ \hline
\end{tabular}%
}
\end{table}

\section{The SoPrA Model: High-level Description}
\label{sec:sopra:highlevel}
The SoPrA model aims to capture the habitual, social and interconnected aspects of SPT. We use the elements as presented in the SPT literature as a basis: actions (or in our new terminology: activities), a meaning element and a physical element. To adequately capture habits, social intelligence and interconnected actions, we use the presented theory from other disciplines. This section presents a high level overview of SoPrA, the next section shows how SoPrA meets the requirements presented in Table \ref{table:overview}.

\begin{figure}[h!]
\center
\includegraphics[width=\textwidth]{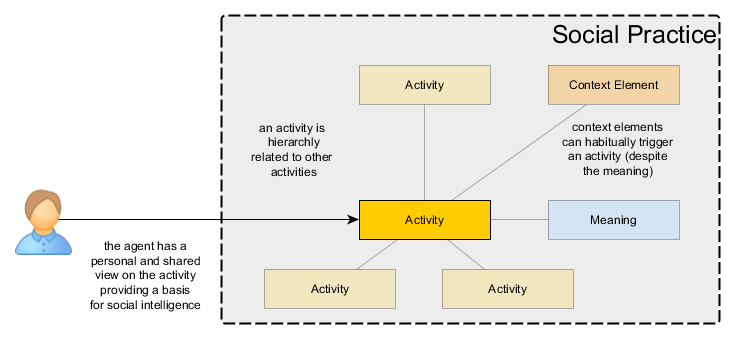}
\caption{An abstract overview of SoPrA depicting how it captures the habitual, social and interconnected aspects of social practices.}
\label{figure:SoPrAOverview}
\end{figure}

Figure \ref{figure:SoPrAOverview} presents a high-level overview of SoPrA. We use activities as the central component of SoPrA, because they are fundamental to each core aspect we aim to model: social practices consists of activities, habits are relations between context and activities, activities are the basis for social cognition and the basis for the activity hierarchies humans use to represent the world.

These activities are firstly used to capture the habitual aspects of SPs by relating the activities to context elements. An activity will be habitually enacted when in the (performance) context of an agent there will be enough context elements that habitually trigger the action. One takes the car to work, because one is at home, just had breakfast and sees the keys on the table. As such, a SP can become a routine: in the same context the same activities, in the same order are performed. 

A routine only gains significance when it can be contrasted with a non-routine mode of acting. For this reason, an activity is associated with a meaning element. The meaning element represent a shared end or purpose for the activity. It provides a reason for the agent to deviate from the routine and intentionally chose an action.

These activities are secondly used to capture the interconnected aspect of SP by relating them to other activities. These activities form a hierarchy that provides a structure that the agent follows when making decisions. To implement one activity one needs to do another activity. After taking the car to work, one needs to take the car to sports. The structure can also be used to make inferences: given that one activity implements the other they will share some properties.

These activities are thirdly used to capture the social aspects of SP by relating them to the beliefs of an agent. Agents have a personal take on an activity and an idea of how others view the activity. These views often overlap. One thinks car driving is efficient, one's partner thinks car driving is efficient and so, without saying, both assume they will carpool to work with the car. Agents base their social intelligence on their shared view on activities. As such, a SP can be used as a shared social world to guide interactions.

A practices truly becomes social when multiple agents have the same view on an activity. What constitutes a \emph{social} practice, instead of a personal practice, can thus only be determined in a bottom-up descriptive fashion. Every practice exists on an continuum ranging from personal to shared. The extent to which a practice is shared changes over time: daily social dynamics influence which views become prominent or dwindle. Moreover, there is no one cut-off point where a practice is shared 'enough' to become social: every interaction requires a different threshold of sharedness. For example, to go through a door requires less of a shared social world then playing soccer. This paper provides groundwork by expressing for every agent in which view they belief. Future work will explore how the dynamics of social interactions are dependent on the level of sharedness a social practice provides.

SoPrA thus associates elements (e.g., meaning, context elements) with these activities and not --- as is common in SPT literature --- with the SP as a whole. For example, an agent does not associate the SP of commuting with environmentalism. Instead, an agent associates an implementation of commuting --- taking the train to work --- with environmentalism, while associating another implementation --- taking the car to work --- with the meaning of efficiency.  We give three advantages of this modelling choice. First, it allows to capture the beliefs an agent has about a SP in a nuanced way (e.g., make a difference between forms of commuting). Second, it allows to use these nuances to guide the decision-making of the agent (e.g., choose car-driving over the train because its more efficient). Third, it allows one to feed the model with a selection of facts about the actions and let the agent make inferences about more complex activities (e.g., commuting is boring, because all of its implementations are boring). This inheritance from activity to activity will get more attention in Section \namerefopt{sec:sopra:formalrules}.



\section{The SoPrA Model: UML Model}
This section provides a detailed overview of SoPrA in the Unified Modelling Language and several modelling choices. The model is presented in parts and linked to the requirements. For the full model see \namerefopt{appendix:UML.full}.
\label{sec:sopra.uml}



\subsection{Activity Tree}
The basic building block of SoPrA are activities. They provide the core of a SP (\ref{req:SP.actions}). These activities connect to each other both temporally and ontologically (\ref{req:IA.ontologicaltemporal}). SoPrA enables one ontological relation known as the generalization. The generalization association allows the agent to make inferences about the properties of an activity based on its parent (\ref{req:IA.inferences}) as will be further discussed in Section \namerefopt{sec:sopra:formalrules}. For pragmatic reasons we simplify the temporal relation to the use of the parthood association. The parthood association specifies that one activity is composed of several other activities. In other words, all the child activities need to be completed to complete the parent activity. The parthood association allows us to model sequential activities. For example, car-commuting consists of bringing the kids to school and going to work. Note that we do not specify the order in which these activities should be done nor if they can be done simultaneously. In terms of \citet{Okeya2014}, we can model sequential activities (\ref{req.IA.sequential}), but not overlapping activities (\ref{req.IA.overlapping}). 
By enabling temporal and ontological associations between activities SoPrA provides an activity hierarchy that agents use to order and represent the world (\ref{req:IA.basicbuildingblock},\ref{req:IA.hierarchy}).

\begin{figure}[h!]
\center
\includegraphics[height=0.15\textheight]{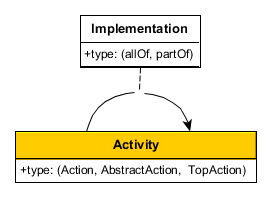}
\caption{An UML diagram depicting the Activity class, the Implementation association class connecting activities and the different types one can attribute to these classes.}
\label{figure:activityclass}
\end{figure}

\begin{figure}[h!]
\includegraphics[width=\textwidth]{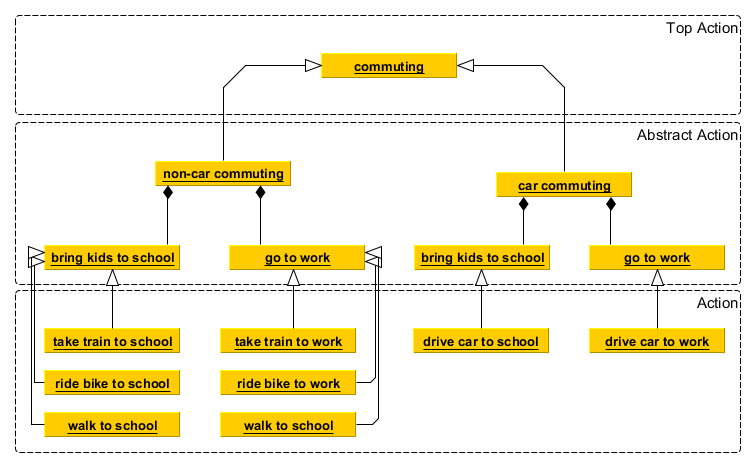}
\caption{An activity tree relating top actions, abstract actions and actions for commuting. The white arrows specify kind-of associations and the black diamonds part-of associations. 
}
\label{figure:spTree}
\end{figure}

Figure \ref{figure:activityclass} presents the \umltext{Activity} class and the \umltext{Implementation} association class. The \umltext{Implementation} association class can have the \umltext{allOf} type, which represents the aforementioned generalization relation, and the \umltext{partOf} type, which represents the aforementioned temporal relation. To easily refer to the right level of abstraction of an activity we differ between three types: actions (atomic), abstract actions and the top action (root). Figure \ref{figure:spTree} depicts an instance of these classes for our use case of commuting. The activity hierarchy should be seen as the basis for the decision-making of the agent. It allows a mechanism where the agent stepwise decides which activity it will do: from the activity of commuting to the activity of taking the car to work. 

\subsection{Activity Associations}
\label{sec:model.activityassociations}
\begin{figure}[h!]
\center
\includegraphics[width=0.75\textwidth]{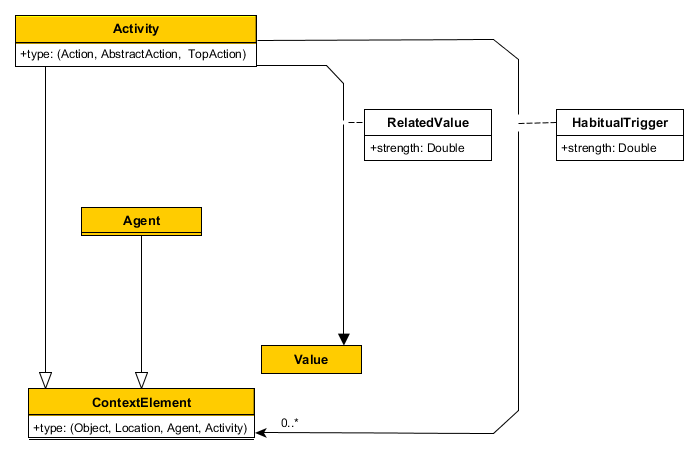}
\caption{An UML diagram depicting activity specific assocations such as the RelatedValue association class and the HabitualTrigger association class.}
\label{figure:UML.activityelements}
\end{figure}

Figure \ref{figure:UML.activityelements} present how we associate a physical element (\ref{req:SPT.context}) and a meaning element (\ref{req:SPT.meaning}) with activities to model SPs as habitual routines (\ref{req:SPT.habits}) that can be contrasted with a slow intentional side (\ref{req:H.automatic}).

SoPrA has a \umltext{HabitualTrigger} association class that associates activities with a \umltext{ContextCue} class. The \umltext{strength} attribute of \umltext{HabitualTrigger} association class represents to what extent the context element is connected with the activity. Habits are thus modelled as continious action-`context cue' associations (\ref{req:H.associations}). These context-cue's include objects, locations, people, other activities and time (\ref{req:H.contextcue}) and are modelled in different ways. We model objects and locations as type of context cue's (i.e., they do not have a seperate class). They represent the physical aspect of a SP (\ref{req:SPT.context}). Agents are modelled as a separate class that specializes (is generalized by) a \umltext{ContextCue}. We reserve a separate class for agents as this class will have associations that other context cue's will not have (as opposed to object and location). Activities are already part of the model and are also a specialization of \umltext{Context Cue}. We do not include time as a possible habitual trigger of activities. Time is a special case as it semantics interplay with other parts of our model (e.g., the parthood association). Due to time and space limitations we postpone the modelling of time as a habitual trigger to modelling the dynamic aspects. The \umltext{HabitualTrigger} association class allows us to model SP as habitually enacted routines (\ref{req:SPT.habits}). 


SoPrA requires a meaning elements that represents a shared end or purpose (\ref{req:SPT.meaning}). We aim to model the concept of meaning as clearly distinguishable, measurable, and understandable in terms of empirical observations (i.e., operationalized) as this serves our aim of creating a validated model of SPT. The literature on SPT provides several candidates that could be used to model meaning: emotions, motives, values, norms or goals.

We argue that the concept of human values (e.g., environmentalism, tradition, achievement) provides a good candidate to model meaning as its (1) teleological, (2) shared and (3) operationalized and (4) trans-situational.
First, values can be defined as `ideals worth pursuing' \citep{Dechesne2012}. Values are ends that people what to achieve: they are unreachable, but can be pursued \citep{Weide2011}. Values thus provide the teleological element we require of our model of meaning. 
Second, \citet{Schwartz2012} developed several instruments (e.g. surveys) to measure and categorize values. Values thus provide an operational concept that has been made measurable and understandeble in terms of empirical observations.
Third, these findings on intervalue comparison have been extensively empirically tested and shown to be consistent across 82 nations representing various age, cultural and religious groups \citep{Schwartz2012,Schwartz2012a,Bilsky2011,Davidov2008,Fontaine2008}. Values thus provide a shared concept to which different people can relate. This satisfies our requirement to use shared social concepts as the basis of agent reasoning (\ref{req:SI.socialbuildingblocks}). 
Fourth, values are trans-situational \citep{Weide2011}. This allows us to interconnect actions from different domains (\ref{req:SPT.interconnected}). For example, environmentalism can relate both to the practice of commuting as well as dining.

Figure \ref{figure:UML.activityelements} shows how we model the connection between values, context and activities. In addition to an activity and a value class, there is an association class called \umltext{RelatedValue}. The \umltext{RelatedValue} association class has a \umltext{strength} attribute with type \umltext{Double} that represents how strongly an activity is attached to a value. Analogue the \umltext{HabitualTrigger} association class has a strength attribute that represents how strongly a context element triggers an activity. 

\subsection{Connecting Agents and Activities}
\begin{figure}[h!]
\center
\includegraphics[height= 0.4\textheight]{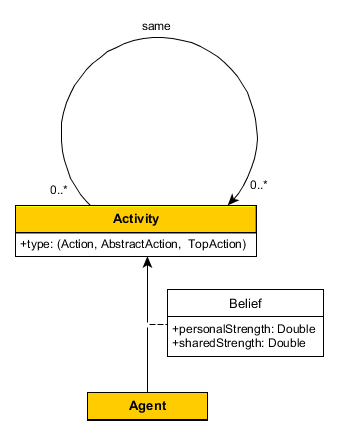}
\caption{An UML diagram presenting how agents and activities are connected with the belief association class. The same association connects activities that represent the same bodily movement.}
\label{figure:agentbelief}
\end{figure}

To enable SPT for agent-based simulations, SoPrA needs to model the connection between SP and an agent (see Figure \ref{figure:agentbelief}).

SoPrA connects agents and activities with the \umltext{Belief} association class. If multiple agents belief in the same activity the SP is similar for them (\ref{req:SPT.socialsimilar}). A SP can then form a common ground between agents that gives them the ability to represent and reason about a shared social world (\ref{req:SI.sharedworld}). The activity class here represents one view on an activity. This allow for multiple views (\ref{req:SI.multipleviews}), that is, multiple instantiations of the activity class for the same bodily movement. For example, \umltext{drive-car-to-school1} represents the view that car-driving is fun, while \umltext{drive-car-to-school2} represents the view that car-driving is not fun.

The \umltext{Belief} association class has an attribute \umltext{personalStrength} that represents to what extent the agent itself agrees with this view (\ref{req:SI.personalview}). The attribute \umltext{sharedStrength} represents to what extent an agent beliefs others agree with this view (\ref{req:SI.othersview}). This allows the agent to have independent beliefs how others view an activity and how itself views this activity (\ref{req:SI.independentviews}). For example, I belief \umltext{drive-car-to-school1} (i.e., car-driving is fun), but most authors belief that \umltext{drive-car-to-school2} (i.e., car-driving is not fun). Note that in a simpler model, one that connects these attributes with the \umltext{Activity} class, these attributes are necessarily shared over multiple agents (e.g., violates \ref{req:SI.independentviews}). 

The \umltext{Activity} class has a reflexive same association that represents that two instances of the activity class (e.g., \umltext{drive-car-to-school1} and \umltext{drive-car-to-school2}) refer to the same bodily-movement. This allows agents to recognize which instances represent different views on the same bodily-movement. This will help agents in making inferences on how different views on an activity connect. For example, Bob wants to enact \umltext{drive-car-to-school1}. He beliefs that most others view the \emph{same} bodily movement as \umltext{drive-car-to-school2}: driving the car to school promotes the value of social interaction. Therefore Bob reasons that his neighbour will agree when he asks him to take his kids to school as well. These inferences thus enable the agent to use the common ground SP provide to represent and reason about the social world (\ref{req:SI.sharedworld}).


\subsection{Agent Associations}
\begin{figure}[h!]
\center
\includegraphics[height= 0.25\textheight]{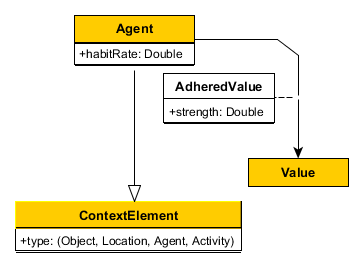}
\caption{An UML diagram depicting agent specific elements such as the habitRate and the AdheredValue assocation classs.}
\label{figure:agentass}
\end{figure}

Figure \ref{figure:agentass} depicts the \umltext{Agent} class with its attributes that are needed to deal correctly with habitual behaviour and contrast this with intentional behaviour (\ref{req:SPT.habits}).

Agents have a attribute that ensures heterogeneity in how they acquire habits. This attribute \umltext{habitRate} represent how much the \umltext{strength} of the \umltext{HabitualTrigger} association class increases when an agent experiences an action in relation to a \umltext{ContextElement}. As shown in \citep{Mercuur2017}, a difference in \umltext{habitRate} suffices to acquire heterogeneity in both the learning rate (\ref{req:H.difflearningtime}) as well as the maximum habit strength the agents report  (\ref{req:H.diffmaximum}). Future work will describe the dynamics of learning habits.

We model an association class between agents and values called \umltext{AdheredValue} with an attribute \umltext{strength}. This represent to what extent an agent values a value. This enables agents to intentionally choose one action over another. For example, an agent might value environmentalism highly and relate the activity \umltext{Non-Car Commuting1} to environmentalism, therefore, it chooses \umltext{Non-Car Commuting1} over \umltext{Car Commuting1}. We require SoPrA to have a basis for intentional actions as habits become only meaningful when contrasted with intentional behaviour (\ref{req:H.automatic}). We chose values to represent the intentional (or teleological) in SoPrA for the reasons we explained in section \ref{sec:model.activityassociations}: values are operationalized, shared and trans-situational.

\subsection{Additional Formal Rules}
\label{sec:sopra:formalrules}
SoPrA contains a number of logical rules that are necessary to correctly depict SPT, that is, fulfill the requirements extracted from the literature (see Table \ref{table:overview}). We view these logical rules in two ways. First, as constrains that limit the number of possible instantiations of SoPrA to those that correctly depict SPT. Second, as reasoning rules for agents to make inferences about the SP. We use first-order predicate logic to describe these rules as it is a well-known logic. In \citet{Mercuur2018DL} we show that these rules can be translated to description logic such that we can use automatic reasoners to enforce the aforementioned constrains and inferences. 

The activity hierarchy is a directed tree. This implies the activity hierarchy has a unique root. This choice is made such that the root can be used as a starting point in the decision-making process. A directed tree can be formally defined as a connected graph where there is one more node then there are edges. Thus, the activity hierarchy $H$ is a directed tree 
$H = < A, r, I>$, where $A$ is the set of instances of the \umltext{Activity} class, with root activity $r\in A$, and where $I : A \times A$ is the set of instances of \umltext{Implementation} associations between activities, $I^+$ is its transitive closure, such that 
\begin{equation}
(\forall a\in A\backslash\{ r\}: 
(a,r)\in I^+)
\wedge
(|I| = |A| - 1)
\end{equation}

An activity can have three different types: action, abstract action and a top action. An action is an atomic leaf in the activity tree, an abstract action is in the middle and a top action is the root node of the tree. These types are defined in SoPrA to allow the agent to have a starting point (top action) and an ending point in the decision-making (action). We can formally define these types as follows:
\begin{equation}
    \forall x \in A (\neg  \exists y \in A \textit{ }I(y,x) \leftrightarrow \textit{ Action}(x))
\end{equation}
\begin{equation}
    \forall x \in A((r = x) \leftrightarrow \textit{ TopAction}(x))
\end{equation}
\begin{equation}
    \forall x \in A (\neg(Action(x) \lor TopAction(x)) \leftrightarrow \textit{ AbstractAction(x)})
\end{equation}

The implementation association can have the \umltext{partOf} type, which should represent that one activity is composed of several other activities. In other words, if an activity implements part of another activity, then there should be another activity that also implements part of this parent activity. We express this formally as,
\begin{equation}
\label{eq:partof}
\forall c_1,p \in A (\textit{ ImplementsPartOf}(c_1,p) 
\rightarrow 
(\exists c_2 \in A \textit{ ImplementsPartOf}(c_2,p) 
\wedge c_1 \neq c_2))
\end{equation}

The activity hierarchy implies a logical relation between the activity associations.\endnote{The inheritance of activity associations implies that there are multiple versions of an activity based on their place in the activity tree. For example, bringing kids to school as part of non-car commuting is different then bringing kids to school as part of car commuting; the first one promotes environmentalism, the last one does not.} For example, if all implementations of non-car commuting (taking a train, bike or walk) are strongly associated with environmentalism, then non-car commuting is also strongly associated with environmentalism. In general, the strength with which the values are associated with a parent activity is the average of the strength with which the values are associated with the children.
 To express this formally we need to define two functions. First the function $C: A \mapsto A^n$  where $n \in \mathbb{N}$, that maps a parent activity $p$ to its set of children $C(p)$
\begin{equation}
C(p) = \{c\in A \mid I(c,p)\}.
\end{equation}

We define the \umltext{RelatedValue} association class such that it is a function that maps the set of values $V$ and activities $A$ to the \umltext{strength} attribute of the \umltext{RelatedValue} association. For all $a\in A$, $v\in V$ let $r(a,v): A \times V\mapsto \mathbb{R}$ denote the strength of the relation between value $v$ and activity $a$. This strength is determined as the average strength of the relation between the same value and each of the children of the activity. If $a$ is an action, then this strength is specified in a seperately defined function $u$. Formally:
\begin{equation}
\label{eq:valueinheritance}
 \forall a\in A, \forall v\in V :
s(a,v) = 
\begin{cases}
    \frac{\sum_{c\in C(a)} s(c,v)}{|C(a)|}& \text{if } C(a)\neq \emptyset\\
    u(a,v)            & \text{otherwise}
\end{cases}
\end{equation}
where $u(a,v): \textit{Action} \times V \mapsto \mathbb{R}$ is a separately defined function that denotes the strength of the relation between $v$ and the actions. This function depends on the domain and use case and can for example be based on empirical data. 

Note that the antecedent $C(p) \neq \emptyset$ ensures $p$ has children it can inherit the values from. We do not enforce any logical rules on the inheritance of the habitual trigger association. Humans can have a habit for a parent activity without having a habit for its children and \emph{vice versa}. Even allowing for two conflicting habits can lead to a theoretically sound model (as shown in \citep{Mercuur2017}).

Recall that SoPrA enables agents to make inferences about which activities refer to the same bodily movement with the \umltext{same} association. The \umltext{same} relation represents an equivalence class. This means that the association is reflexive, symmetric and transitive.

\section{Implementation \& Verification}
\label{sec:implver}

\subsection{Implementation}
We implemented SoPrA in Netlogo \citep{Netlogo} and Prot\'eg\'e \citep{Protege}. The code is available online.\endnote{For a Netlogo implementation for our trafic choice use case, see: https://github.com/rmercuur/SoPraTraffic-Netlogo. For the Prot\'eg\'e implementation, see: https://github.com/PCSan/SOPRA.} 

\begin{figure}
    \centering
    \includegraphics[width=0.5\textwidth]{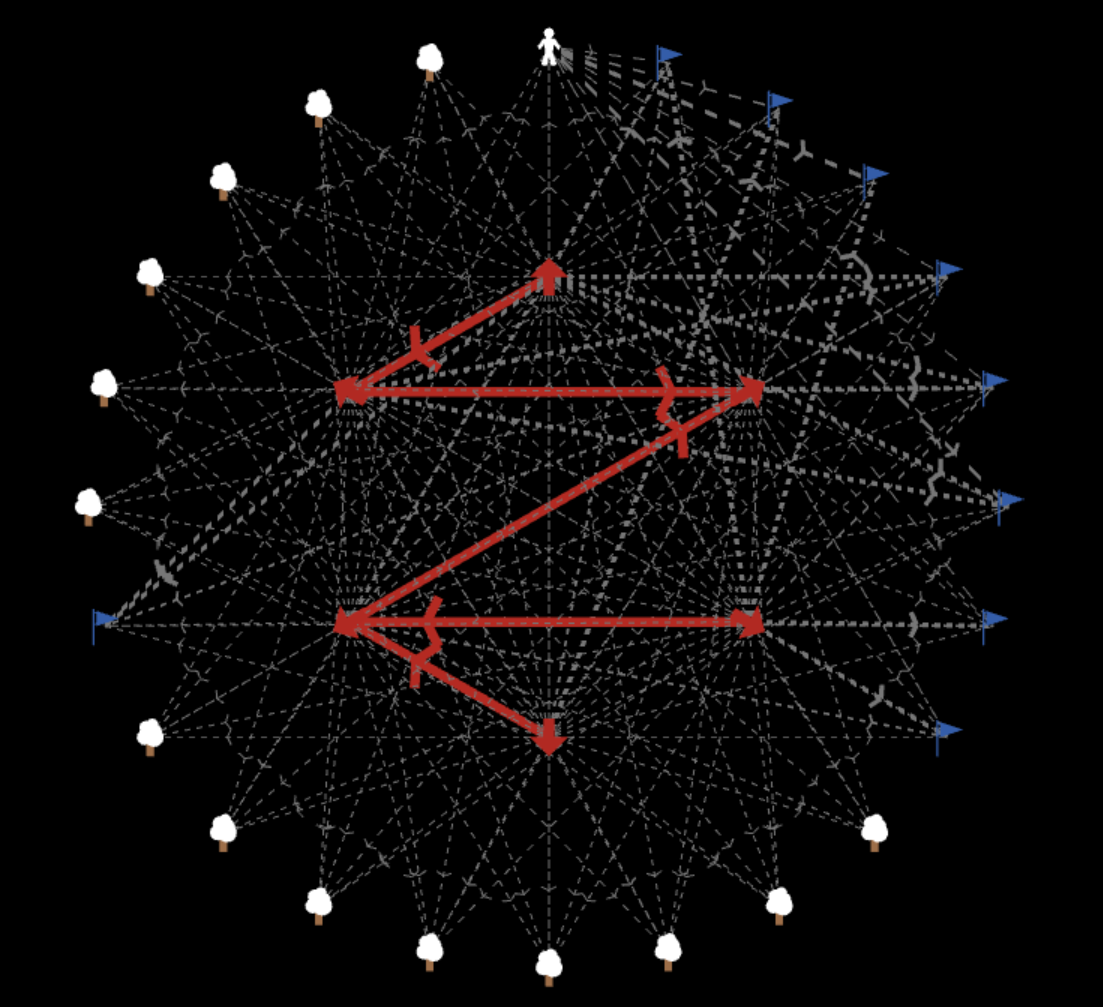}
    \caption{An instantation of SoPrA in Netlogo for one agent that has one view on commuting. The red arrows represent activities and the red lines implementation associations. The circle of white tree's and blue flags represent respectively context elements and values. The remainder of the links represent beliefs, related value, habitual trigger, or adhered value associations.}
    \label{fig:netlogo}
\end{figure}

\begin{figure}[h!]
\center
\includegraphics[width=\textwidth]{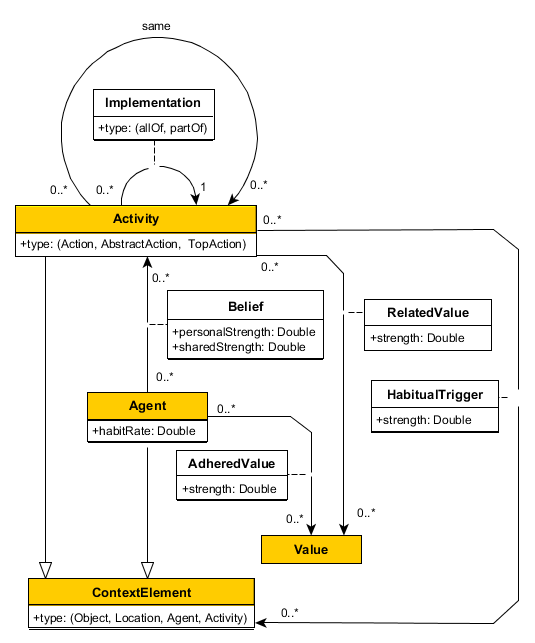}
\caption{A Complete Overview of SoPrA in UML.}
\label{figure:UML.full}
\end{figure}

An implementation of SoPrA in Netlogo allows researchers to infer new knowledge over time. This paper provides the static groundwork for these simulation. Netlogo has been chosen as our agent-based simulation platform for its accessibility for researchers interested in modelling social phenomena. The UML classes are implemented as breeds in Netlogo and associations are links. Links can have variables in Netlogo corresponding to the attributes of association classes in UML. Netlogo does not have built-in mechanisms to represent an object hierarchy (such as presented in the UML in Figure \ref{figure:UML.full}) or formal rules (such as represent in Section \namerefopt{sec:sopra:formalrules}). However, these relations can be captured in manual procedures. Figure \ref{fig:netlogo} depicts a visualization of SoPrA in Netlogo. In future work such visualization can be used to visualize how SPs and links between them evolve over time.

The implementation of SoPrA in Prot\'eg\'e allows researchers to author formal ontologies and enable formal reasoners that aid the modeller. First, the formal reasoners can infer new static knowledge. For example, one can use the formal reasoner to apply formal rule (\ref{eq:valueinheritance}) and automatically infer that if riding a bike, taking the train to work and walking promote environmentalism then non-car commuting also promotes environmentalism. Second, it can check if SoPrA is satisfiable. For example, if the formal rules of Section \ref{sec:sopra:formalrules} would be inconsistent Prot\'eg\'e would detect and report this. Third, the formal reasoner can check if SoPrA is properly instantiated. For example, if one would assert that an activity is partly implemented by only \emph{one} child then Prot\'eg\'e would report this is inconsistent with rule (\ref{eq:partof}).

Prot\'eg\'e makes this possible by its use of description logic: a logic that trades its expressivity to acquire decidability \citep{Horrocks2007}. The UML classes correspond to classes in Prot\'eg\'e and associations to object properties. In \citet{Mercuur2018DL}, we show how to capture association classes and the formal rules in Prot\'eg\'e. Consequently, this shows that description logic is expressive enough to capture the semantics of SoPrA. 



\subsection{Verification}
Verification is the act of testing whether the computational model does what it is supposed to be doing \citep{Grabner2018}. This amounts to testing if it adequately implements the underlying conceptual model. In our case, we want test if the conceptual model in Section \namerefopt{sec:sopra.uml} is adequately implemented by (1) the formal model in Prot\'eg\'e and (2) the computational model in Netlogo.

The two implementations are cross-checked. The conceptual model in Netlogo can be translated to Prot\'eg\'e models using the method described in \citet{polhill2015}. A Netlogo model should be satisfiable and properly instantiated in Prot\'eg\'e. If the models hold up this means either both models are correct implementations or both models are wrong in exactly the same way. A positive result thus lower the chance of one of the two implementations being wrong.

In Netlogo we run unit tests. We check if certain properties of the conceptual model that are not directly implemented still hold. For example, the amount of belief associations should not exceed the amount of agents multiplied by the amount of activities. A positive result lowers the chance that the implementation has properties that do not represent the conceptual model.

Prot\'eg\'e can check if the model is satisfiable and correctly instantiated. Clearly we belief the conceptual model is satisfiable and correctly instantiated. It thus checks if the implementation has properties we belief the conceptual model has. A positive results implies a higher chance that the model is correctly implemented or that there is something wrong with the conceptual model. Prot\'eg\'e automatically generates explanations for possible inconsistencies that can help to detect which of the two is the case.




\section{Applying SoPrA to our Use Case}
\label{sec:usecase2}
This section presents how SoPrA can be used to capture the habitual, social and interconnected aspects of our use case on CO2-emission due to car commuting. We specify the assignment of SoPrA, which part of the model should be mapped on which part of the target \citep{Weisberg2013}. The assignment of SoPrA is a necessary step to do validation \citep{Grabner2018}. Given that this paper focus on the static part, we map states of the model on the target (and not for example, mechanism). It is always easier to validate aspects of a model that are a direct and explicit representation of real-world objects \citep{Schulze2017}. This section demonstrates there is a direct and explicit mapping of SoPrA to the habitual, social and interconnected aspects of the scenario in Section \namerefopt{sec:usecase}. 

We capture the scenario in terms of SoPrA in Table \ref{tab:scenario}. We go through the scenario step by step and explain how SoPrA represents this step in Table \ref{tab:scenario}.

\setlist[description]{style=unboxed}
\begin{description*}
\item[Bob values the environment] Line 28 shows that Bob adheres to the value of enviromentalism with a high strength.
\item[Bob has a habit of taking the car to bring the kids to school and go to work] Line 25-27 show that there is HabitualTrigger relation between Kid1 and these activities. Note how SoPrA forces one to specify that it is specifically the kid that triggers the habit. Furthermore, line 16, 18 and 20 show how Bob personally beliefs in this view on these activities. 
\item[Alice asks Bob if he wants to go by train with her.] This is a consequence of the following statements.
\begin{description*}
\item[Alice beliefs Bob wants to commute] Line 13 shows that Alice beliefs that \umltext{Commuting1} is a view on commuting that is shared. Line 24 shows that \umltext{Commuting1} is an activity that is usually triggered at home. Thus she infers that Bob (who is at home) would like to commute.
\item[Alice beliefs that people that go by car usually value comfort] Line 15 and 34 show that Alice beliefs that \umltext{Car Commuting1} is usually seen as an implementation of \umltext{Commuting 1}. Line 29 shows that \umltext{Commuting1} promotes the value of comfort. Alice thus infers that Bob who usually does \umltext{Car Commuting1} is looking for an implementation of \umltext{Commuting1} that promotes comfort.
\item[Alice sees the rain outside and thinks Bob will find riding a bike uncomfortable, but the train comfortable] Line 31 and 32 show that the views Alice beliefs to be shared on these activities state that the train promotes comfort more than the bike.
\item[Alice beliefs that normally when others bring the kids to school by car, they will also go to work by car] Line 15, 17, 19, 36 and 37 show that Alice beliefs that \umltext{Bringing Kids to School1} is sharedly seen as an implementation of \umltext{Car Commuting1} and that this activity has another part: \umltext{Going to Work with the Car1}. 
\item[However, this situation is different] Alice beliefs that the current situation could call for another activity as there is no need for the activity \umltext{Bring Kids To School 1}. Line 22 shows that Alice beliefs (both personally and wrt to others) in another implementation of Commuting 1. Namely, one with only one part named \umltext{Going To Work 1} (i.e., there is no bringing the kids to school). As \umltext{Going To Work 1} is an implementation of \umltext{Non-Car Commuting1}, she asks Bob to reconsider his travel mode.
\end{description*}
\item[Bob agrees to go by train, because it promotes environmentalism and he does not have a habit in this situation] There are no context-elements that could trigger \umltext{Car-Commuting 1} over \umltext{Non-Car Commuting 1} (or vice versa). Line 33 shows that \umltext{Go To Work with Train 1} promotes environmentalism.
\end{description*}

\begin{table}[]
\centering
\caption{A table showing a possible model that describes our use case. Only relevant instances are mentioned (e.g., possible \umltext{RelatedValue} and \umltext{Habitualtrigger} that do not play a role in the scenario are omitted.)}.
\label{tab:scenario}
\resizebox{\textwidth}{!}{%
\begin{tabular}{llllll}
\hline
\textbf{Number} & \textbf{Class} & \textbf{Instance 1} & \textbf{Instance 2} & \textbf{Attribute 1} & \textbf{Attribute 2} \\ \hline
1 & Activity & Commuting 1 &  & type: TopAction &  \\
2 & Activity & Car Commuting 1 &  & type: AbstractAction &  \\
3 & Activity & Bring Kids To School With Car 1 &  & type: Action &  \\
4 & Activity & Go to Work with Car 1 &  & type: Action &  \\
5 & Activity & Non-Car Commuting 1 &  & type: AbstractAction &  \\
6 & Activity & Go To Work 1 &  & type: AbstractAction &  \\
7 & Activity & Go To Work With Train 1 &  & type: Action &  \\
8 & Activity & Go To Work With Bike 1 &  & type: Action &  \\
9 & Agent & Bob &  & habitRate: 0.8 &  \\
10 & Agent & Alice &  & habitRate: 0.5 &  \\
11 & Agent & Kid1 &  & habitRate: 0.8 &  \\
12 & ContextElement & Home &  & type: location &  \\
13 & Belief & Alice & Commuting 1 & personalStrength: 0.1 & sharedstrength: 0.8 \\
14 & Belief & Bob & Commuting 1 & personalStrength: 1.0 & sharedStrength: 0.8 \\
15 & Belief & Alice & Car Commuting 1 & personalStrength: 0.1 & sharedstrength: 0.8 \\
16 & Belief & Bob & Car Commuting 1 & personalStrength: 1.0 & sharedStrength: 0.8 \\
17 & Belief & Alice & Bring Kids To School With Car 1 & personalStrength: 0.1 & sharedstrength: 0.8 \\
18 & Belief & Bob & Bring Kids To School With Car 1 & personalStrength: 1.0 & sharedStrength: 0.8 \\
19 & Belief & Alice & Go to Work with Car 1 & personalStrength: 0.1 & sharedstrength: 0.8 \\
20 & Belief & Bob & Go to Work with Car 1 & personalStrength: 1.0 & sharedStrength: 0.8 \\
21 & Belief & Alice & Non-Car Commuting 1 & personalStrength: 1.0 & sharedStrength: 0.8 \\
22 & Belief & Alice & Go to Work 1 & personalStrength: 1.0 & sharedStrength: 0.8 \\
23 & Belief & Alice & Go to Work With Train 1 & personalStrength: 1.0 & sharedStrength: 0.8 \\
24 & HabitualTrigger & Commuting 1 & Home & strength: 1.0 &  \\
25 & HabitualTrigger & Car Commuting 1 & Kid1 & strength: 1.0 &  \\
26 & HabitualTrigger & Bring Kids To School With Car 1 & Kid1 & strength: 1.0 &  \\
27 & HabitualTrigger & Go to Work with Car 1 & Kid1 & strength: 1.0 &  \\
28 & AdheredValue & Bob & Environment & strength: 0.8 &  \\
29 & RelatedValue & Commuting 1 & Comfort & strength: 0.9 &  \\
30 & RelatedValue & Car Commuting 1 & Comfort & strength: 0.9 &  \\
31 & RelatedValue & Go To Work With Train 1 & Comfort & strength: 0.7 &  \\
32 & RelatedValue & Go To Work With Bike 1 & Comfort & strength: 0.0 &  \\
33 & RelatedValue & Go To Work With Train 1 & Environment & strength: 1.0 &  \\
34 & Implementation & Car Commuting 1 & Commuting 1 & type: allOf &  \\
35 & Implementation & Non-Car Commuting 1 & Commuting 1 & type: allOf &  \\
36 & Implementation & Bring Kids To School With Car 1 & Car Commuting 1 & type: partOf &  \\
37 & Implementation & Go to Work with Car 1 & Car Commuting 1 & type: partOf &  \\
38 & Implementation & Go To Work 1 & Non-Car Commuting 1 & type: allOf &  \\
39 & Implementation & Go To Work With Train 1 & Go to Work 1 & type: allOf &  \\
40 & Implementation & Go To Work With Bike 1 & Go to Work 1 & type: allOf &  \\ \hline
\end{tabular}%
}
\end{table}

\section{Conclusion}
This paper presented the SoPrA model that enables the use of Social Practice Theory for agent based simulations. SPT provided new insights in sociology, but to show that SPT can provide potential explanations through simulation we need a computational model that is validated. This paper provides the groundwork with a computational model that is a correct depiction of SPT, computational feasible and can be directly mapped to the habitual, social and interconnected aspects of a target scenario.

Section \namerefopt{sec:literature} extracted requirements from the literature that summarize the core aspects of SPT. We made the choice to focus on habits, social intelligence and interconnected actions and therefore dropping elements such as competences and affordances that were not directly instrumental to these core aspects. To correctly model these core aspects we used supporting literature from social psychology, neuroscience and agent theory.
Section \namerefopt{sec:sopra.uml} presented an UML model (added with formal rules) that fulfills these requirements. We made the choice to model the temporal relations between action as the parthood association such that we can model sequential activities, but not overlapping activities. Section \namerefopt{sec:implver} presented publicly available implementations in Netlogo and Prot\'eg\'e that can be cross-checked to verify a proper implementation of the UML. Section \namerefopt{sec:usecase2} showed how one can model a scenario regarding the social practice of commuting. The section demonstrated how one maps the model on a system and how SoPrA can capture statements that combine habits, social intelligence and interconnected element. For example, it can capture how Alice understands that Bob makes an exception to join her in the train commute, because he doesn't have to bring the school. We thus showed how SoPrA is a correct depiction of SPT through extracting and fulfilling requirements, the computational feasibility through presenting a model in UML and making a verified implementation in Netlogo and Prot\'eg\'e and a clear mapping from target to phenomena through a use case of commuting.

The next step is to extend the model with dynamic mechanisms that can generate output from input and validate this extended model against empirical data. By validating the model one can show that SoPrA can provide potential explanations and increase our understanding of social phenomena. Future work aims to explore possibilities to use SoPrA outside agent-based simulations to guide human-agent interaction in socially aptitude and effecient way and to create transparant and explainable AI.

\newpage
\section{Appendix A: A Complete Overview of SoPrA in UML}
\label{appendix:UML.full}



\theendnotes


 
\bibliographystyle{jasss}
\bibliography{Mendeley} 


\end{document}